\documentclass{article} 
\begin{document} 
\begin{center} 
\title{Amplitude for existence of spacetime points}
\author{M.Dance} 
\end{center} 

\begin{abstract} 
This paper outlines a possibility for spacetime dynamics and structure, without postulating a metric ab initio. In this model, the closer an object is to a mass or energy source, the more paths through spacetime might be available to the object in the direction of the mass/energy, or the higher the amplitude associated with such paths. A simple possibility might be that spacetime points $x$ have an amplitude of existence $E(x)$ consistent with this.  The magnitude of $E(x)$ might be greater or less than 1 at any point $x$; or the relative values of $E(x)$ might be what matters. In a classical limit, a function like $E(x)$ might give the effect of a gravitational metric. 
\end{abstract} 

\section{Introduction}

Quantum gravity is an evolving field with many streams of ongoing work, and there is room for entirely new theories. An early concept  was Snyder's proposal of a quantum spacetime with non-commuting coordinates \cite{snyder 1947}. There are numerous good reviews of quantum gravity, e.g. \cite{rovelli 06}, \cite{kiefer 05}, \cite{isham 95} and \cite{isham 99}. 

Logically, there are many possibilities for a theory of quantum gravity.  As noted by \cite{isham 95}, one possibility is that the theory might require a fixed set of spacetime points with a topology and/or differential structure on this set that is subject to quantum effects. 

This paper covers a simple suggestion that I made in private correspondence during the mid-1990s. This paper elaborates the suggestion and brings it to a wider audience, in case it might be useful. 

\section{A perspective on spacetime}

Let us visualise an object orbiting a planet. Without presupposing a gravitational metric, let us take a step back and ask what might give the effect of a metric. In a quantum mechanical (or similar) picture which sums over possible paths, the bending of the object's motion towards the planet due to gravity might be described by postulating that the closer the object is to the planet, the more paths through spacetime are fundamentally available to the object, or the higher the amplitude associated with such paths.  In this scenario, I postulate that the amplitude of any given path (from purely spacetime factors) need not be simply a phase factor with unit magnitude.  In the classical limit, a metric could be derived from the amplitudes. 

There are a number of potential options at the outset for a path-based model. 

A simple possibility might be that spacetime points $x$ have an amplitude of existence $E(x)$.  The function $E(x)$ need not be restricted to unit amplitude. As I originally envisaged it, $E(x)$ would represent a {\em probability} amplitude that the point $x$ exists, so the magnitude of $E(x)$ would be between 0 and 1.  However, a path integral picture (or consideration of Green's functions) involving products of $E(x)$ for series of spacetime points might imply instead that $|E(x)|$ should generally need to be close to 1; it might be infinitesimally greater than 1 in some regions, and less than 1 in other regions of spacetime. Variation must be the case for an $E(x)$ function to underpin non-flat spacetimes.  Alternatively, it may only be the relative values of the $E(x)$ that really matter; perhaps they need not be centred around 1. 

One might consider $E(x)$ to be a "propensity" amplitude - see e.g. \cite{sorkin 95} for a discussion of propensity.  Hartle has also suggested that "virtual probabilities" outside [0,1] might be useful as intermediate steps in analyses of probabilities \cite{hartle 04}.  From a standard quantum mechanical standpoint, $|E(x)| \neq 1$ need not cause non-unitary evolution of fields on the spacetime, as I would expect the relevant total wavefunctions of objects/fields to be appropriately normalizable. 

One could construct various types of models for the dynamics of a function like $E(x)$.  For example, the closer $x$ is to mass/energy, the larger $|E(x)|$ might be. This type of model may give rise to a circularity: $E(x)$ would be dependent on distance in some way, while (1) distance would be defined by the metric, (2) the metric would be derived from amplitudes, and (3) the amplitudes would depend on $E(x)$. This circularity may present an opportunity to find self-consistent solutions.

One might think that $E(x)$ would be just another scalar field, and one might ask what it adds. However, this would not be an ordinary scalar field theory. $E(x)$ would not enter a Lagrangian in the typical scalar field form added to a standard pre-existing Lagrangian term for gravity. Instead, $E(x)$ might modify the measure of integration and might alter the form of derivatives throughout the theory, or might require other additional terms if derivatives cannot be defined.  $E(x)$ would not itself be a quantum mechanical wavefunction. In contrast to approaches involving a wavefunction of the universe, the path integral would not integrate over manifolds; rather, $E(x)$ would be uniquely a solution of the dynamical equations resulting from the theory.

I present this description despite the seeming likelihood that no observer will ever be able to observe down to a spacetime "point". (Under present theories, it appears likely that distances cannot be observed more closely than the Planck length.)  Notwithstanding the observational difficulties, an $E(x)$ theory based on a continuum spacetime might be useful.  Currently Lagrangians are generally constructed using fields on a spacetime continuum. Non-commutative geometry also exists on a spacetime continuum, while also implying a minimum observable distance interval and eliminating many of the divergences in standard quantum field theory. That is, it seems that a minimum observable distance interval need not require a discrete underlying spacetime.

In the presence of a probability amplitude such as $E(x)$, an observer who assumes (after consideration or otherwise) that $E(x) = 1$ for all $x$ might simply arrive at a different coordinate system on that assumption. Any intervals with $E(x) = 0$ might simply not be noticed as missing, and the coordinates might just be relabelled. This would presumably lead to a corresponding transformation of the observed metric. However, it seems a priori possible that an $E(x)$ formulation might offer advantages.

A spacetime probability amplitude like $E(x)$ might effectively provide a kind of random landscape on which matter fields would propagate. As a speculative conjecture, such a spacetime might even appear dynamically 4-dimensional, or might provide an arrow of time due to an increased density of paths or increased $|E(x)|$ in what human observers interpret as being a timelike direction. 

It might be possible to formulate a theory which combines a factor like $E(x)$ with non-commutative geometry.

\section{Reduction to classical mechanics}

According to the usual prescriptions, classical mechanics requires that trajectories make the action stationary. Taking the limit as $\hbar$ approaches zero in any path integral (e.g. for any particular form of $E(x)$) gives that classical requirement, and the classical solutions emerge. 

With an $E(x)$ function, it is possible that an effective action could have a non-zero imaginary part. This would reflect an exponential decay of the amplitude for a path to explore more regions with $|E(x)| < 1$ than absolutely necessary.  That is, such paths would be exponentially suppressed. Taking the limit as $\hbar$ approaches zero would presumably have the effect of requiring the classical dynamics to take the most highly weighted path, all "forces" considered, including an effective metric which might emerge in the classical limit.  On a simple view, it then seems that a function like $E(x)$ could  possibly underpin classical general relativity. 

It may be appropriate to revisit the basis for using Lagrangians and actions in the light of an amplitude such as $E(x)$.  The action and the Lagrangian are somewhat odd concepts. They are based on Newtonian mechanics.  Penrose has commented recently about the Lagrangian approach \cite{Penrose 04} that:

\small{"...I confess my unease with this as a fundamental approach....In most situations, the Lagrangian density does not itself seem to have clear physical meaning; moreover, there tend to be many different Lagrangians leading to the same field equations.... I remain uneasy about relying upon them too strongly in our searches for improved fundamental physical theories. ..." }

\section{Ordinary quantum mechanics as limit in flat spacetime}

In standard quantum mechanics, a wavefunction $\psi(x)$ is usually interpreted as the probability amplitude for the particle to be at the point $x$. There are a variety of opinions on how quantum mechanics or variants should be formulated and interpreted, and I assume the standard approach. With an $E(x)$ function, an amplitude of a particle being at $x$ might be given by $\psi(x) E(x)$. The amplitude $\psi(x)$ might be reinterpreted as a conditional probability amplitude: the probability amplitude that the particle is at $x$, assuming that the point x exists.  

It would not be sufficient to simply replace $\psi(x)$ in all equations with $\psi(x) E(x)$; as noted above, integration measures and derivatives would presumably depend on $E(x)$, or there may be additional terms.  The overall quantum mechanical wavefunction would also need to be normalized by a factor which would depend on $E(x)$.

It is to be expected that in a spacetime that is flat, $E(x)$ might be near 1 everywhere, subject to fluctuations. If so, only paths with $|E(x)|$ near 1 throughout the path would contribute significantly. In the hypothetical case in which $|E(x)|$ is 1 for all $x$, derivatives (and/or other terms) and measures would presumably revert to their usual form and the theory would be expected to revert to standard quantum theory.

As noted above, an $E(x)$ function would be substantially different in nature and dynamics from a scalar field. It seems unclear whether $E(x)$ could be quantised in the usual sense, or whether it could have a ground state and excited states. If it can, one could define operators that annihilate and create excitations, in a manner likely to depend on mass/energy.  If an $E(x)$ ground state has non-zero energy, standard quantum theory would not quite be achieved, and additional vacuum energy would be present (dark matter?). In that case, the $E(x)$ function might be able to create excited states of itself in a positive feedback process.

\end{document}